\documentclass[twocolumn,twoside,slac_two]{revtex4}

\usepackage{graphicx}
\usepackage{xspace}
\usepackage{fancyhdr}

\newcommand\Swift{\textsl{Swift}\xspace}

\newcommand\Fermi{\textsl{Fermi}\xspace}

\newcommand\fdg{\mbox{$.\!\!^\circ$}}%

\begin{document}

\title{TANAMI counterparts to IceCube high-energy neutrino events}

\author{F. Krau\ss{}$^{1,2}$, B. Wang$^{1,2,3}$, C. Baxter$^{1,2,4}$,
  M. Kadler$^{2}$, K. Mannheim$^{2}$, R. Ojha$^{5,6,7}$, C.
  Gr\"afe$^{1,2}$,
  C.~M\"uller$^{2,1}$, J. Wilms$^{1}$, B. Carpenter$^{5,6}$, R.
  Schulz$^{1,2}$\\
 on behalf of the TANAMI and the \Fermi-LAT Collaborations
\vspace*{0.2cm}}

\affiliation{$^{1}$Dr. Remeis Observatory/ECAP, FAU, Sternwartstr.7, 96049
  Bamberg, Germany}
\affiliation{$^{2}$Institut f\"ur Theoretische Physik und Astrophysik,
  Universit\"at W\"urzburg, Emil-Fischer-Str. 31, 97074 W\"urzburg, Germany}
\affiliation{$^{3}$Department of Physics and Astronomy, University of Pittsburgh, Pittsburgh, PA 15260 USA}
\affiliation{$^{4}$School of Physics and Astronomy, University of
  Edinburgh, Old College, South Bridge, Edinburgh, EH8 9YL, Scotland}

\affiliation{$^{5}$NASA, Goddard Space Flight Center, Greenbelt, MD 20771, USA}
\affiliation{$^{6}$Catholic University of America, Washington, DC 20064, USA}
\affiliation{$^{7}$University of Maryland, Baltimore County,
  Baltimore, MD 21250, USA}

\begin{abstract}
  Since the discovery of a neutrino flux in excess of the atmospheric
  background by the IceCube Collaboration, searches for the
  astrophysical sources have been ongoing.
  Due to the steeply falling background towards higher energies, the
  PeV events detected in three years of IceCube data are the most
  likely ones to be of extraterrestrial origin.
  Even excluding the PeV events detected so far, the neutrino flux is
  well above the atmospheric background, so it is likely that a number
  of sub-PeV events originate from the same astrophysical sources that
  produce the PeV events.
  We study the high-energy properties of AGN that are
  positionally coincident with the neutrino events from three years of
  IceCube data and show the results for event number 4. IC 4 is a
  event with a low angular error ($7\fdg1$) and a large deposited
  energy of 165\,TeV.
  We use multiwavelength data, including \Fermi-LAT and X-ray data,
  to construct broadband spectra and present parametrizations of the
  broadband spectral energy distributions with logarithmic parabolas.
  Assuming the X-ray to $\gamma$-ray emission in blazars originates in
  the photoproduction of pions by accelerated protons, their predicted
  neutrino luminosity can be estimated. The measurements of the
  diffuse extragalactic background by \Fermi-LAT gives us an estimate
  of the flux contributions from faint unresolved blazars. Their
  contribution increases the number of expected events by a factor of
  $\sim$2. We conclude that the  detection of the IceCube neutrinos
  IC4, IC14, and IC20 can be explained by the integral emission of
  blazars, even though no individual source yields a sufficient energy
  output.  
\end{abstract}

\maketitle

\thispagestyle{fancy}

\section{INTRODUCTION}

The IceCube Collaboration's announcement of the discovery of a
neutrino flux in excess of the atmospheric background is an inflection
point in multimessenger astronomy \citep{IceCube2013b}.
Due to the steeply falling atmospheric background spectrum, events at
the highest energies most likely have an extraterrestrial origin
\citep{IceCube3}. 

Neutrino emission from the jets of active galactic nuclei (AGN)
\citep{Mannheim1995} and cores \citep{Stecker2013} has been predicted,
but alternative possibilities 
are gamma-ray bursts \citep{Waxman1997} and pevatrons in the Galactic
center region \citep{Aharonian1996}. All IceCube events are consistent
with an isotropic distribution, and therefore extragalactic sources
are the prime candidates.
Only the predicted flux of
$\sim10^{-8}$\,GeV/cm$^2$/s/sr at energies from 100\,TeV to a few PeV
from AGN jets matches the observed excess flux well \citep{Learned2000}.

AGN with jets that are observed at small angles to the line of sight are
called `blazars'. Their non-thermal emission becomes relativistically
boosted. The low energy emission is generally attributed to
synchrotron emission. Emission at higher energies can be explained by
hadronic and leptonic models.
In hadronic models, protons (as well as electrons) are accelerated in
the jet. The protons interact with seed photons at lower energies
(e.g., from the accretion disk or external radiation fields) and
produce pions \citep[pion photoproduction; ][]{Mannheim1989}. Subsequent
pion decays produce neutrinos and $\gamma$-rays.
Currently, the observed spectral energy distributions (SEDs) of AGN can be
described equally well with hadronic and leptonic emission processes due to a
large number of free parameters \citep[e.g, ][]{Boettcher2013}.
Unambiguous evidence of hadronic processes could be provided by an
association of neutrino events with an individual blazar. In pion
photoproduction, the neutrino flux can be directly calculated from the
observed flux of the high-energy bump in the SED $F_\nu = F_\gamma$.
This estimate has been confirmed by Monte-Carlo simulations
\citep{Muecke2000}. The neutrino fluence can therefore be estimated
directly from the integrated X-ray to $\gamma$-ray flux of the
broadband SED.

Due to the large angular uncertainties, several possible candidate
blazars can be identified for each of the IceCube shower events. We have
previously shown \citep{eb} that the 2 events at PeV energies from the
first two years of IceCube (IC20, dubbed `Ernie' and IC 14,
`Bert') can be explained calorimetrically by the six candidate blazars
from the TANAMI sample. 
Here,we study the multiwavelength properties of AGN from the TANAMI
sample, as well as \Fermi blazars that are positionally coincident with
the neutrino events from three years IceCube data. We address the
question whether the sub-PeV neutrino events can be explained by
blazars in the error field. In particular, we calculate the expected
neutrino fluence of the the four blazars in the field of IceCube event
4 (IC4).
IceCube event number 4 has a lower median angular error of $7\fdg1$
compared to the PeV events with error radii of up to $13^\circ$ and a
higher energy than most of the other IC events (165\,TeV), i.e., has a
low probability of being of atmospheric origin. Inside the IC4 error
field, there are four $\gamma$-ray bright AGN listed in the 2LAC
catalog \citep{2lac}. We report on the multiwavelength properties of
these four sources below.

\section{MULTIWAVELENGTH DATA}

Tracking Active Galactic Nuclei with Austral Milliarcsecond
Interferometry
(TANAMI)\footnote{\url{http://pulsar.sternwarte.uni-erlangen.de/tanami}}
\citep{Ojha2010} is a multiwavelength program that monitors
extragalactic jets of the Southern Sky.
 
\begin{figure}
\includegraphics[width=0.4\textwidth]{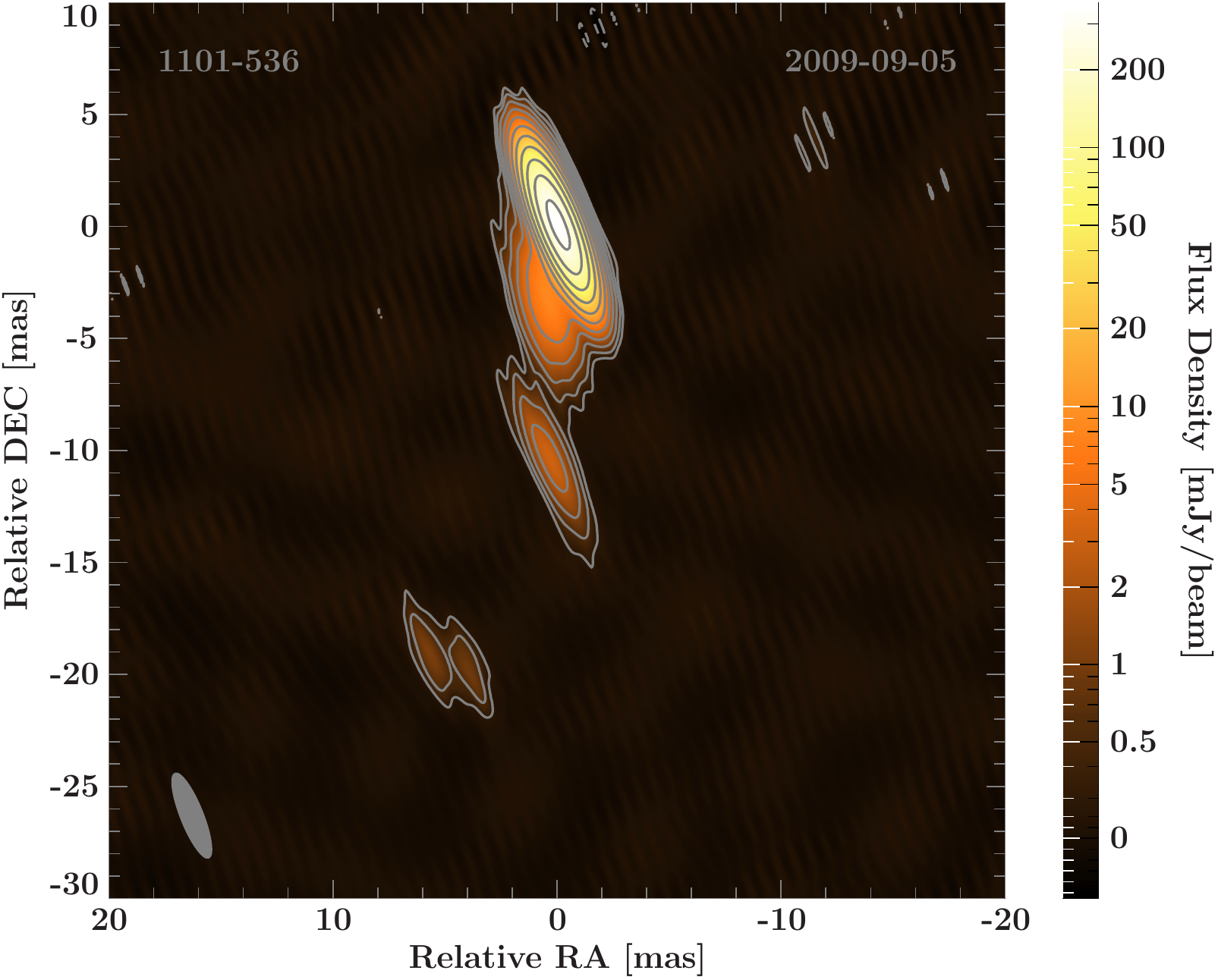}
\caption{First-epoch TANAMI VLBI image of 2FGL\,J1103.9$-$5356 at
  8.4\,GHz. The color scale indicates the flux density distribution,
  the white contours are scaled logarithmically and increase by
  factors of 2, with the lowest contour set to the
  3$\sigma$-noise-level. The gray ellipse in the lower left corner
  shows the beam with $(4.1\times 1)\,$mas at $22^\circ$. This blazar
  shows a bright radio core with a brightness temperature of
  $T_\mathrm{B}=5.43\times10^{10}$\,K (for
  $S_\mathrm{core}\sim0.39$\,Jy) and a single-sided jet in southern
  direction.  
  \label{plot1101}
} 
\end{figure}

Figure~\ref{plot1101} shows the first-epoch high-resolution image of
2FGL\,J1103.9$-$5356 (PKS\,1101$-$536) obtained with Very Long Baseline
Interferometry (VLBI) at 8.4\,GHz. An 8.4\,GHz VLBI image of
PKS\,1104$-$445 has been shown by \cite{Ojha2010}.
Both sources show core-dominated radio structures typical for blazars
with a single-sided jet, indicating relativistically beamed emission.
The two other IC\,4 candidate sources have not been observed in the
TANAMI VLBI program as of 2015.

X-ray data taken during the IceCube period are from the
TANAMI program and the public archives of Swift \citep{Swift2004} and
\textsl{Chandra}. 
\Swift/XRT and \textsl{Chandra}/ACIS data were reduced with standard
methods, using the most recent software packages (HEASOFT 6.15.1, CIAO
4.6) and calibration databases. 
Spectra were grouped to a minimum signal-to-noise ratio of 5 to ensure
the validity of $\chi^2$ statistics. For a low SNR, the
spectra were grouped to a minimum signal-to-noise ratio of 2 and the
use of Cash statistics \citep{Cash}.
Spectral fitting was performed with ISIS 1.6.2 \citep{ISIS2000}.
The X-ray data were deabsorbed using
the Galactic $N_\mathrm{H}$ value \citep{Kalberla}, abundances from
\cite{Wilms2000}, and cross sections from \cite{Verner1996}. 
We have used the $\gamma$-ray spectra from the 2FGL catalog
\citep{2fgl}.

\section{RESULTS}

Electromagnetic cascades in pion photoproduction emit at X-ray and
$\gamma$-ray energies, and we approximate the non-thermal photon flux
$F_\gamma$ by the integrated flux between 1 keV and 5 GeV \citep{eb}.
The broadband spectra were fit with a logarithmic parabola
\citep{Massaro2004} including X-ray absorption.

\begin{figure}
  \includegraphics[width=0.49\textwidth]{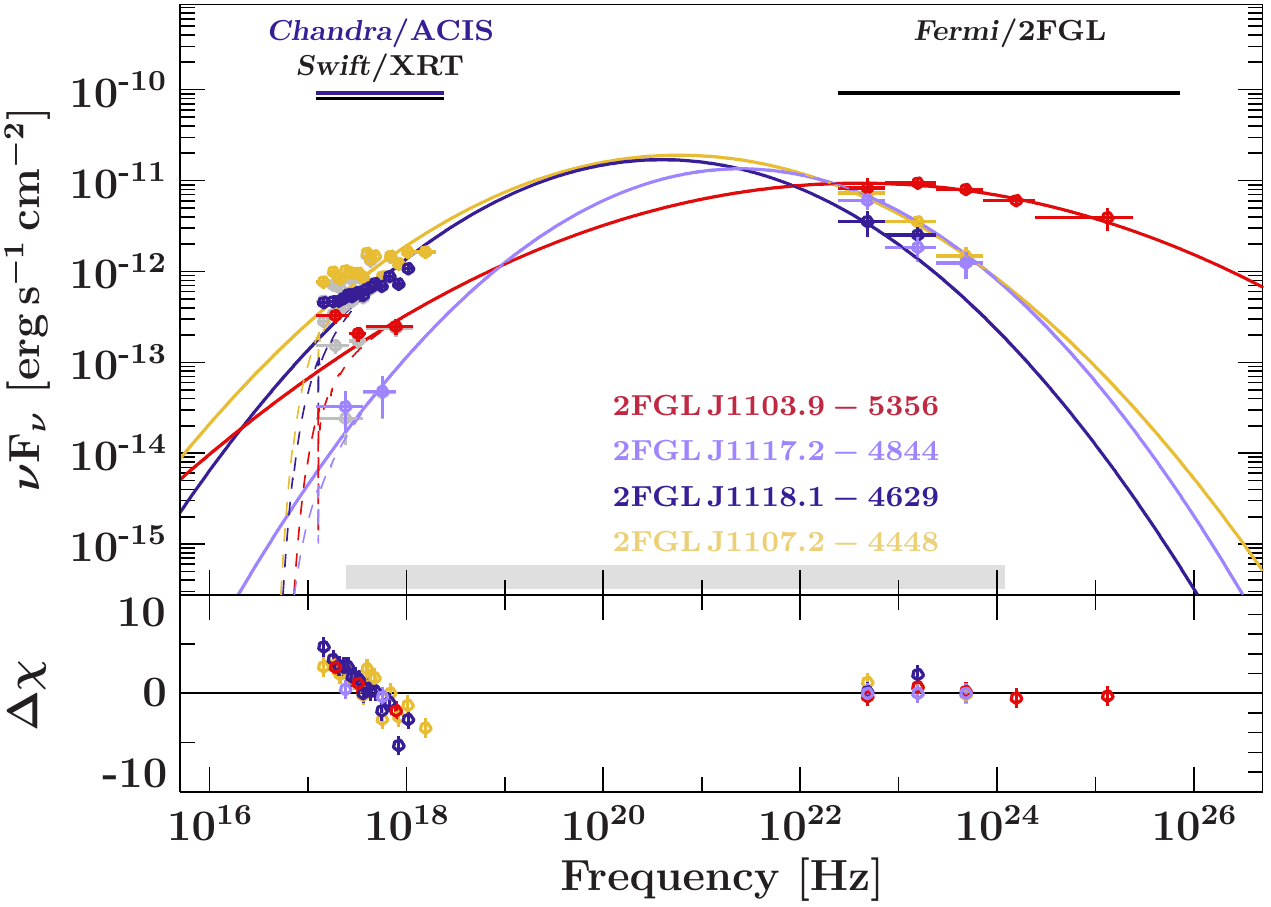}
  \caption{X-ray to $\gamma$-ray SED of all four 2LAC sources
    including a log parabola fit to the data. The gray area shows the
    energy range used for the calculation of the neutrino events}
  \label{sed1}
\end{figure}
The X-ray to $\gamma$-ray SEDs of all four sources are shown in Fig.
\ref{sed1}. 
As shown by \cite{eb}, this allows us to model the high-energy hump with
logarithmic parabolas in order to estimate the integrated flux and the
fluence in the IceCube integration period. This fluence can be used to
directly estimate the number of neutrinos.
Using the IceCube integration period of $\Delta t = 998$ days, and an
effective area of $A_{\mathrm{eff}} = 10^5\,\mathrm{cm}^2$ for
contained events, we obtain the values listed in Table \ref{tabres}.
The numbers would be lower for a realistic spectrum of the emitted
neutrinos or if some fraction of the emission is produced in a
leptonic, proton-synchrotron, or Bethe-Heitler process.
The steepness of the blazar $\gamma$-ray luminosity function
\citep{Singal2012} further implies that in a large field, the neutrino
fluence will have significant contributions from the brightest sources
in the field, as well as from fainter, unresolved sources.

\begin{table}
  \caption{Integrated electromagnetic energy flux from 1 keV to 5 GeV
    and expected electron neutrino events in 998 days of IceCube
    data for the 4 candidate blazars of IceCube event 4. Uncertainties are
    statistical only.} 
  \label{tabres}
  \begin{tabular}{lp{2.6cm}ll}
    \hline
    Source & Assoc. source & $F_\gamma\,[10^{-11}]$&events \\
    2FGL & &[erg/s/cm$^2$]&\\
    \hline
    J1103.9$-$5356 & PKS\,1101$-$536 &  $7.6^{+1.7}_{-1.4}$ &$0.22\pm0.05$\\
    J1107.2$-$4448 & PKS\,1104$-$445 & $14.0^{+1.7}_{-1.8}$ & $0.40^{+0.05}_{-0.06}$ \\
    J1117.2$-$4844 & PMN\,J1117$-$4838 & $8^{+6}_{-5}$ & $0.23^{+0.18}_{-0.15}$\\
    J1118.1$-$4629 & PKS\,1116$-$46 & $11.3\pm0.6$ & $0.33\pm0.02$\\
    \hline
    \textbf{Sum} & & &$\mathbf{1.18\pm 0.18}$\\
    \hline
  \end{tabular}
\end{table}

\subsection{Contributions from unresolved blazars}
At the sensitivity of current catalogs, a large number of faint
blazars are not resolved into individual point sources by \Fermi-LAT,
but do contribute to the diffuse extragalactic gamma-ray background
(EGB). In order to calculate the number of expected neutrinos, one
should also consider the substantial contribution of this most
numerous part of the blazar population. 
The fraction of blazars in the EGB has been estimate to lie between
50\% and 80\% \citep{egb50}. 
At 100 GeV, half of the EGB has been resolved into individual blazars (mainly BL
Lac type objects) by \Fermi-LAT \citep{egb}.

We compare the values of the EGB flux to
the total flux of resolved blazars, in order to estimate the
contributions from unresolved blazars, assuming pion photoproduction.
We find a total integrated flux 
for all four 2LAC sources of $F_{100\,\mathrm{MeV}-820\,\mathrm{GeV}}=1.71\times 
10^{-7}\,\mathrm{ph}/\mathrm{s}/\mathrm{cm}^2$, which corresponds to
$3.54\times10^{-6}$\,ph/s/cm$^2$/sr for a 7.1$^\circ$ error field.
The extragalactic background is
$F_{100\,\mathrm{MeV}-820\,\mathrm{GeV}}=7.2\pm0.6\times
10^{-6}$\,ph/cm$^{2}$/s/sr \citep{egb}, a factor of $\sim 2$ higher
than the value for the resolved blazars. This suggest that a
substantial fraction of the extragalactic neutrino flux in this field
originates from faint, unresolved blazars, instead of the bright,
low-redshift sources. 

\section{CONCLUSION}
Assuming that the high-energy emission originates in pion
photoproduction, the maximum expected number of electron neutrino
events from all four 2LAC sources for IC4 is $1.18\pm0.18$
for 998 days. 
This is close to the number of detected events, but given the
different factors that might reduce the neutrino output below the rate
predicted by our basic model (leptonic contributions, neutrino
spectra, etc.; see \cite{eb}) it seems unlikely that any of the
individual brightest blazars in the field of IC4 can explain the
observed neutrino flux. This situation is similar to the fields of the
two PeV neutrinos IC14 and IC21 \citep{eb}, where the predicted
neutrino flux of the six brightest blazars matched the IceCube
observed flux, but the individual sources fell short of yielding
sufficient fluence.
The integral flux of bright individual blazars and faint remote
sources, however, rises a factor of ~2 above the observed flux in this
field, consistent with the hypothesis that the population of blazars
as a whole can explain the IceCube results.

\bigskip
\begin{acknowledgments}
We acknowledge support and partial funding by the Deutsche
Forschungsgemeinschaft grant WI 1860-10/1 (TANAMI) and GRK 1147,
Deutsches Zentrum f\"ur Luft- und Raumfahrt grants 50 OR 1311/50 OR
1103, and the Helmholtz Alliance for Astroparticle Physics (HAP). We
thank J.E. Davis and T. Johnson for the development of the slxfig
module and the SED scripts that have been used to prepare the figures
in this work. This research has made use of a collection of ISIS
scripts provided by the Dr. Karl Remeis-Observatory, Bamberg, Germany
at \url{http://www.sternwarte.uni-erlangen.de/isis/}. The Long
Baseline Array and Australia Telescope Compact Array are part of the
Australia Telescope National Facility, which is funded by the
Commonwealth of Australia for operation as a National Facility managed
by CSIRO. The \Fermi-LAT Collaboration acknowledges support for LAT
development, operation and data analysis from NASA and DOE (United
States), CEA/Irfu and IN2P3/CNRS (France), ASI and INFN (Italy), MEXT,
KEK, and JAXA (Japan), and the K.A. Wallenberg Foundation, the Swedish
Research Council, and the National Space Board (Sweden). Science
analysis support in the operations phase  from INAF (Italy) and CNES
(France) is also gratefully acknowledged. 
\end{acknowledgments}

\end{document}